\documentclass[aps,prx,twocolumn,superscriptaddress, longbibliography]{revtex4-2}

\usepackage{graphicx,xcolor}
\usepackage{amsmath,amsfonts,amssymb}

\usepackage[]{siunitx}

\usepackage{hyperref}

\newcommand{\beginsupplement}{%
        \setcounter{table}{0}
        \renewcommand{\thetable}{S\arabic{table}}%
        \setcounter{figure}{0}
        \renewcommand{\thefigure}{S\arabic{figure}}%
     }

\begin{document}

\title{Surface-near domain engineering in multi-domain x-cut lithium niobate tantalate mixed crystals}


\author{Laura Bollmers}
\email[Author to whom correspondence should be addressed: ]{Laura.Bollmers@uni-paderborn.de}
\affiliation{Paderborn University, Institute for Photonic Quantum Systems (PhoQS), Warburger Str. 100, 33098 Paderborn, Germany}
\affiliation{Paderborn University, Integrated Quantum Optics, Warburger Str. 100, 33098 Paderborn, Germany}
\author{Tobias Babai-Hemati}
\affiliation{Paderborn University, Integrated Quantum Optics, Warburger Str. 100, 33098 Paderborn, Germany}
\author{Boris Koppitz}
\affiliation{Institut für Angewandte Physik, Technische Universität Dresden, 01062 Dresden, Germany}
\author{Christof Eigner}
\affiliation{Paderborn University, Institute for Photonic Quantum Systems (PhoQS), Warburger Str. 100, 33098 Paderborn, Germany}
\author{Laura Padberg}
\affiliation{Paderborn University, Institute for Photonic Quantum Systems (PhoQS), Warburger Str. 100, 33098 Paderborn, Germany}
\affiliation{Paderborn University, Integrated Quantum Optics, Warburger Str. 100, 33098 Paderborn, Germany}
\author{Michael Rüsing}
\affiliation{Paderborn University, Institute for Photonic Quantum Systems (PhoQS), Warburger Str. 100, 33098 Paderborn, Germany}
\affiliation{Paderborn University, Integrated Quantum Optics, Warburger Str. 100, 33098 Paderborn, Germany}
\author{Lukas M. Eng}
\affiliation{Institut für Angewandte Physik, Technische Universität Dresden, 01062 Dresden, Germany}
\affiliation{ct.qmat: Dresden-Würzburg Cluster of Excellence—EXC 2147, TU Dresden, 01062 Dresden, Germany}
\author{Christine Silberhorn}
\affiliation{Paderborn University, Institute for Photonic Quantum Systems (PhoQS), Warburger Str. 100, 33098 Paderborn, Germany}
\affiliation{Paderborn University, Integrated Quantum Optics, Warburger Str. 100, 33098 Paderborn, Germany}

\date{\today}

\begin{abstract}
Lithium niobate and lithium tantalate are among the most widespread materials for nonlinear, integrated photonics. Mixed crystals with arbitrary Nb-Ta ratios provide an additional degree of freedom to tune materials properties, such as the birefringence, but also leverage the advantages of the singular compounds, for example, by combining the thermal stability of lithium tantalate with the larger nonlinear or piezoelectric constants of lithium niobate. Periodic poling allows to achieve phase-matching independent of waveguide geometry and is therefore one of the commonly used methods in integrated nonlinear optics. For mixed crystals periodic poling has been challenging so far due to the lack of homogeneous, mono-domain crystals, which severely inhibit domain growth and nucleation. In this work we investigate surface-near ($< 1$~$\mu$m depth) domain inversion on x-cut lithium niobate tantalate mixed crystals via electric field poling and lithographically structured electrodes. We find that naturally occurring head-to-head or tail-to-tail domain walls in the as-grown crystal inhibit domain inversion at a larger scale. However, periodic poling is possible, if the gap size between the poling electrodes is of the same order of magnitude or smaller than the average size of naturally occurring domains. This work provides the basis for the nonlinear optical application of lithium niobate tantalate mixed crystals.
\end{abstract}

\maketitle

Integrated quantum photonics exploits (second-order) nonlinear optical effects for many applications ranging from single photon generation, filtering, signal analysis, amplification or as interconnects between quantum systems operating in different wavelength regimes \cite{Eckstein2011,Saravi2021,Boes2023,Luo2019,Stefszky2023,Harder2013}. Lithium niobate (LiNbO$_3$, LNO) and lithium tantalate (LiTaO$_3$, LTO) are widely used for nonlinear optical applications due to their large nonlinear optical coefficients, easy availability of large, optical-quality wafers in form of bulk crystals or thin films, as well as a largely developed technology, which can be leveraged for fabrication \cite{ruesing2019IEEE,Weis85,Hum2007QPMReview,Boes2023}. 

Efficient nonlinear optical processes require phase matching between the interacting beams, which can be achieved independently of the material dispersion via quasi-phase matching in periodically-poled ferroelectric domain structures in LNO or LTO \cite{Byer1997,Hum2007QPMReview,Zhao2023}. Mixed crystals of lithium niobate and lithium tantalate (LiNb$_{1-x}$Ta$_x$O$_3$, LNT$_x$) can be created with arbitrary Nb/Ta ratios over the full compositional range\cite{bartasyte2019,XUE2000,Sugii1976,Bashir2023,Roshchupkin2023,Rusing2016LNT}. They allow to interpolate between the refractive indices and dispersions of LNO and LTO, because LNO is negatively birefringent at room temperature, while LTO is positively birefringent. Thus, for a given temperature a crystal composition can be found, where the birefringence vanishes, while the crystal retains its ferroelectric polarization and nonlinearity \cite{Wood08}. Furthermore, LNT$_x$ mixed crystals enable the combination of favorable properties of the individual materials for designing compounds with properties beyond the limitations of LNO and LTO. Examples are combining the superior  thermal stability of LTO, with the larger nonlinear and piezo-electrical coefficients of LNO \cite{Roshchupkin2023,Yakhnevych2023,Bashir2023} or the possibility to create waveguides in layered hetero-structures \cite{Sugii1976,bartasyte2019}.

The engineering of domains in LNTx has been challenging, because the so-far available crystals are \emph{single-crystalline} but \emph{multidomain}: Here, a single crystal is grown above the ferroelectric Curie temperature i.e. in the paraelectric phase, but during cool-down through the paralectric-ferroelectric phase transition, natural domains form randomly in a three-dimensional pattern \cite{Bashir2023,Roshchupkin2023,Koppitz2024}. Recently, it was observed that the as-grown domain pattern in LNT features non-charged 180\textdegree~domain walls, which are parallel to the c-axis, but also so-called head-to-head and tail-to-tail domain wall configurations, which are oriented 90\textdegree~with respect to the spontaneous polarization direction  \cite{Roshchupkin2023}. Such head-to-head or tail-to-tail domain walls are known to be highly charged \cite{Kampfe2014SHG,Singh2022}.

\begin{figure}
	\centering
	\includegraphics[width=1\linewidth]{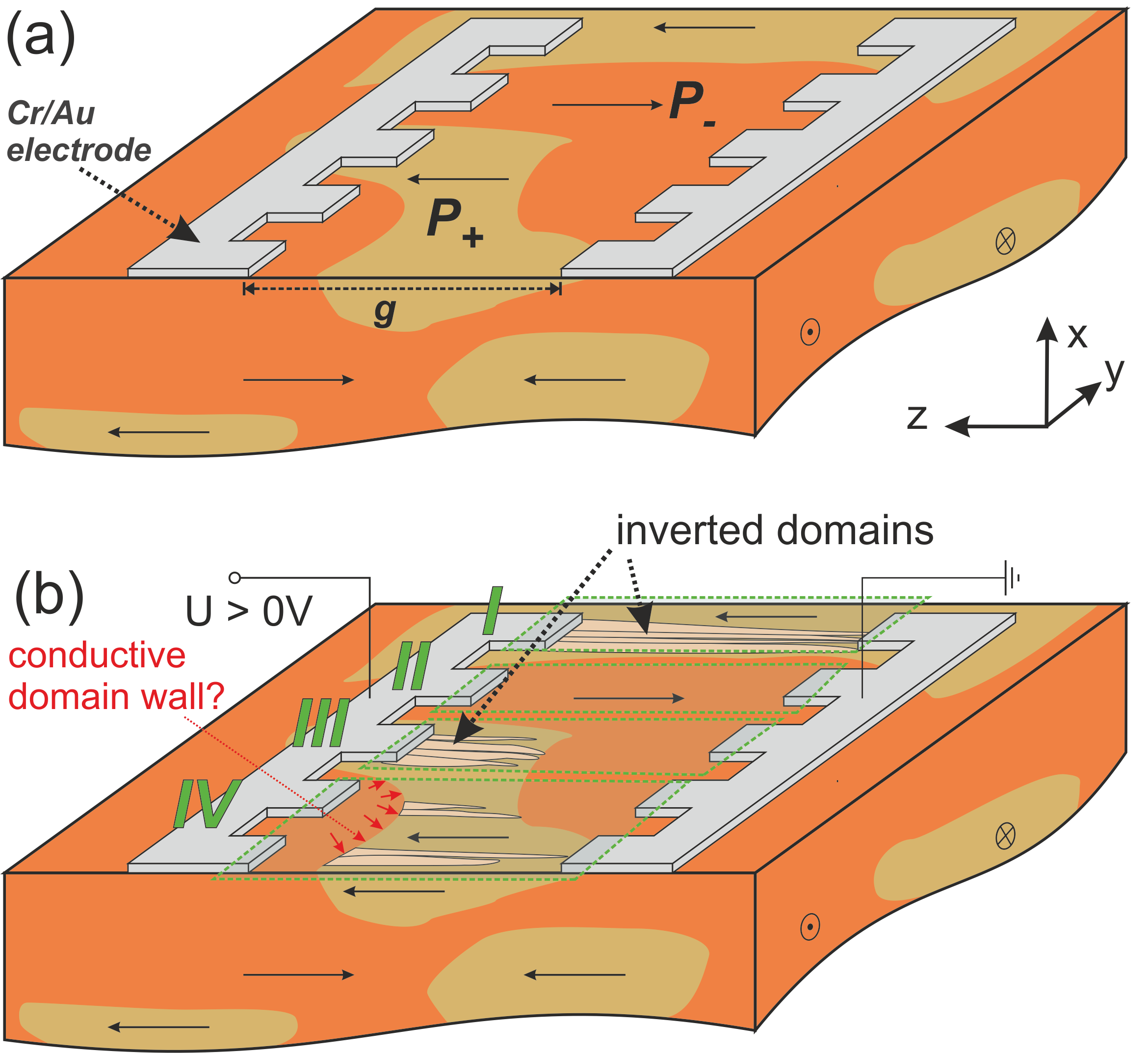}
	\caption{(a) Sketch of multidomain LNT$_x$ crystal: Metal electrodes with a gap $g$ are fabricated on the x-surface of a polished LNT$_x$ crystal. (b) When a positive voltage is applied to the left electrode, four distinct \emph{Cases} of domain inversion and original domain configuration can be distinguished.}
	\label{fig:experimental_geometry}
\end{figure}

The LNT$_x$ system is a uniaxial ferroelectric, i.e. only two domain orientations are possible. They are aligned along the z-axis (crystallographic c-axis). Figure \ref{fig:experimental_geometry}(a) shows a sketch of a random domain pattern in a Czochralski-grown single crystal of LNT$_x$. The average width along the z-axis of the as-grown domains for our sample are in the range of a few 10~$\mu$m. It is suspected that local variations of the Ta/Nb ratio in the range of a few percent points, which naturally occur during growth \cite{Bashir2023}, act as pinning-planes and create very stable as-grown domain patterns, which present a challenge for poling \cite{Koppitz2024}. 

Since it is challenging to produce large, single-domain crystals \cite{Koppitz2024,Roshchupkin2023} , the poling of LNT$_x$ crystals at a local scale of a few 10~$\mu$m is an alternative. This is on the same order or smaller than the original domain sizes of as-grown crystals. For applications in integrated nonlinear optics the cross-section of a periodically-poled area with a width and depth similar to the waveguide cross section is sufficient to achieve quasi-phase matching, because typical waveguides have at most a width of a few microns in both thin film or bulk \cite{Rusing2019SHG,ruesing2019IEEE,Boes2023,Zhao2023}.

In this work we investigate surface-near periodic poling in x-cut multi-domain LNT$_x$ mixed crystals by using lithographically structured metal electrodes across a gap $g$ and the application of an electric field poling at room temperature. Based on the experimental results, we deduce a strategy for periodic poling of such multi-domain crystals. By applying multiple electrical pulses with inverted polarity during the poling process, also a periodic poling can be achieved, even in an originally randomly-patterned domain structure. The crystal has been grown at the Leibniz Institut for Kristallzüchtung (IKZ), Berlin \cite{Bashir2023}. The crystal was oriented via x-ray diffraction, while the Nb/Ta concentration $x$ was confirmed with x-ray fluorescence spectroscopy. For our crystal, we have determined a Ta-content of $x = 0.45 \pm 0.03$. Details on the growth of the crystal and the fundamental characterization can be found elsewhere \cite{Bashir2023}.

Figure \ref{fig:experimental_geometry}(a)  shows the principle experimental geometry for the poling geometry. We deposited  chromium-gold finger electrodes (170~nm chromium and 30~nm gold similar to Stanicki et al. \cite{Stanicki2020}) with a periodic pattern of 10~$\mu$m in y-direction. Multiple electrodes of 250~$\mu$m length and gap distances $g$ ranging from 20 to 30~$\mu$m were fabricated on different regions of the crystal.  The overall poling procedure and electrode design is similar to the widely studied poling of x-cut thin film or bulk LNO \cite{Mackwitz2016,Younesi2021,Gui2009,ZHANG2022,Wang2018}. 

\begin{figure}
	\centering
	\includegraphics[width=1\linewidth]{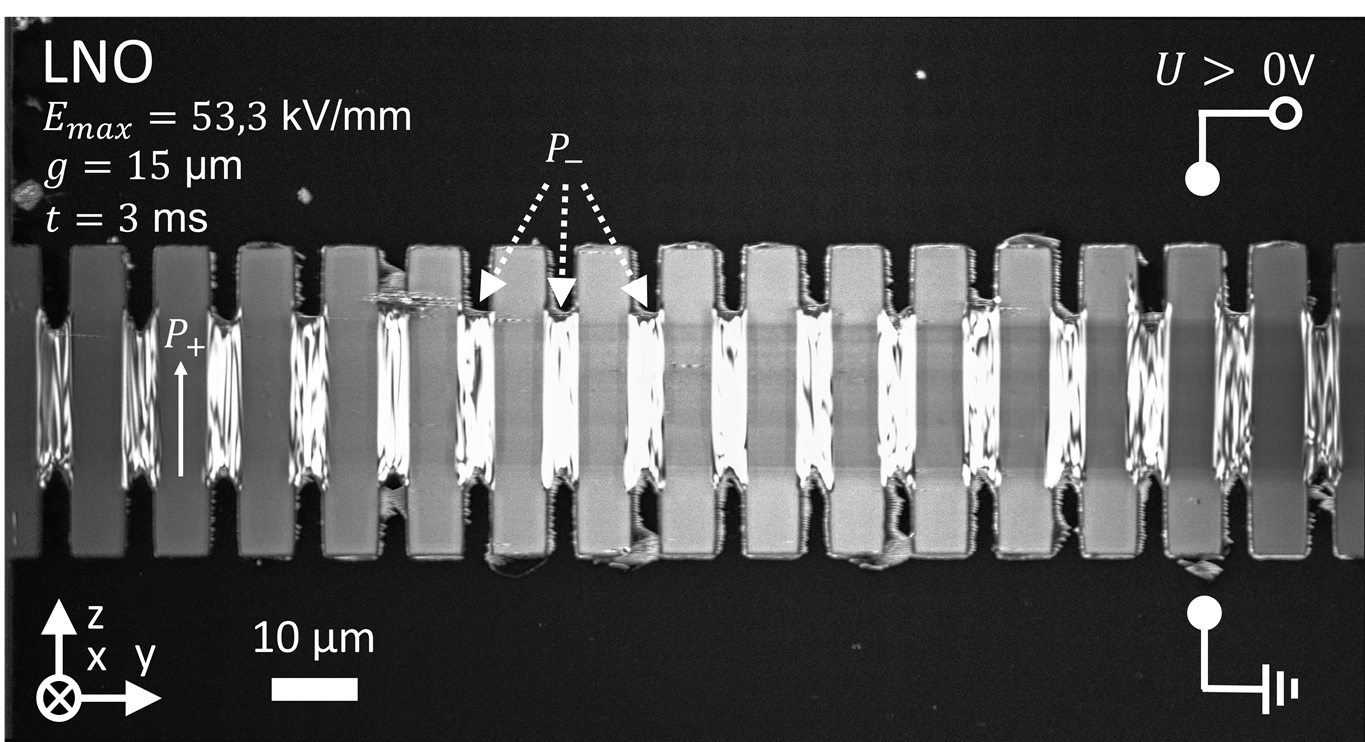}
	\caption{Second-Harmonic microscope image of a poling result on single-domain, x-cut congruent lithium niobate. After application of a single voltage pulse, surface-near domain inversion can be achieved. The results, both poling and SHM contrast, are very similar to previous works \cite{Gui2009,Stanicki2020,Younesi2021}.} 
	\label{fig:cLN}
\end{figure}

\begin{figure*}
	\centering
	\includegraphics[width=1\linewidth]{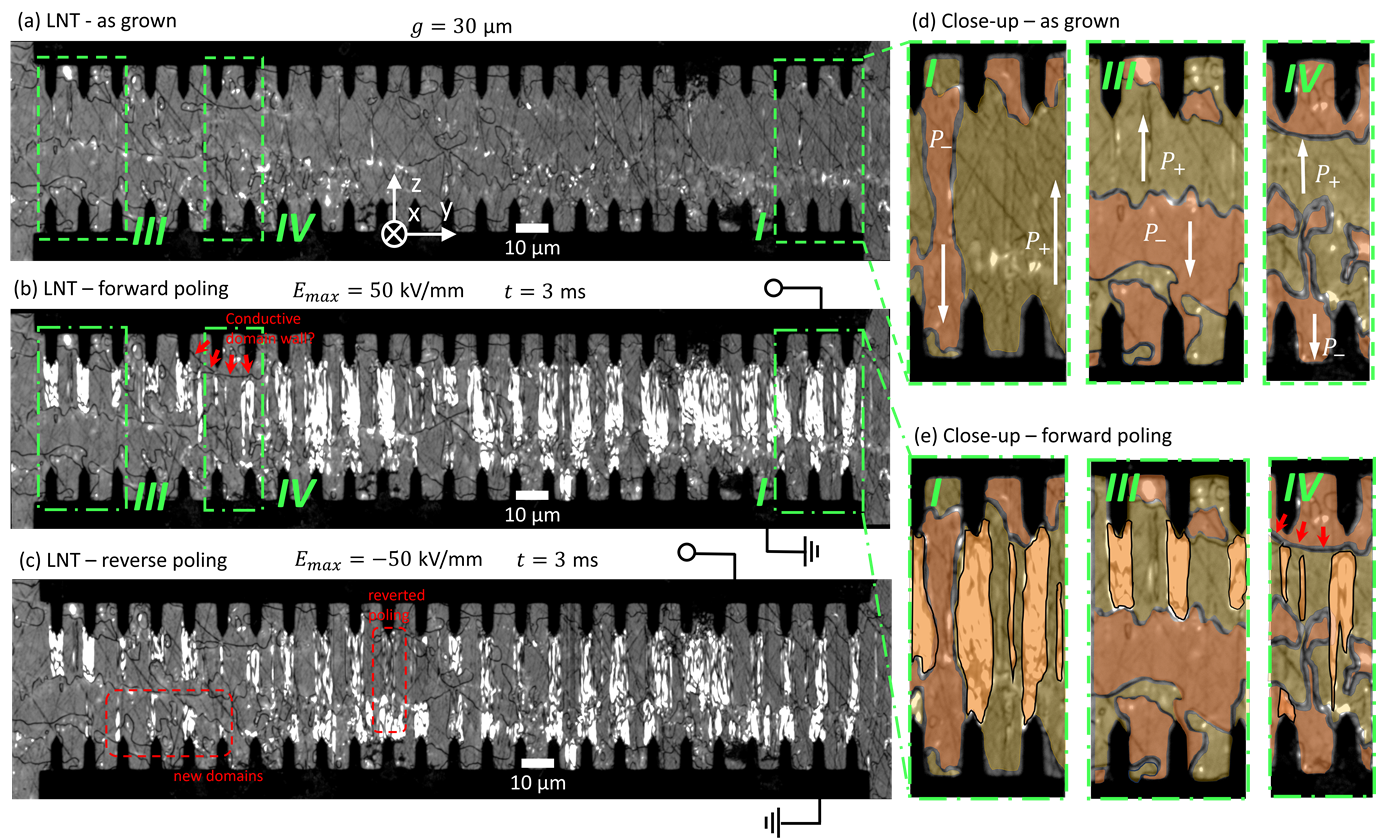}
	\caption{Second-Harmonic microscope image of a 250~$\mu$m-long electrode pair with a gap of $g=30$ ~$\mu$m on x-cut lithium niobate tantalate mixed crystals (a) before poling, (b) after application of a positive voltage pulse with an electric field of $E_{max} = 50$~kV/mm and a duration of $t=3$~ms and (c) after application of the same electric pulse with reversed polarity. (a) Before poling, a random domain structure with meandering domains of 10 to 20~$\mu$m sizes is visible. (b) After forward poling signatures for surface-near, shallow domains with a depth of $0.5-1$~$\mu$m can be seen in agreement with results on congruent x-cut lithium niobate \cite{Mackwitz2016,Younesi2021,Gui2009,ZHANG2022,Stanicki2020}. (c) After application of a reverse pulse, inverted domains also nucleate at the bottom electrodes. Due to apparent strong pinning only little back poling is observed and a quasi-periodic structure can be achieved. Further, close-ups from special regions (d) before and (e) after the positive poling pulse are shown. Here, the interpretation of the original and inverted domain structure is indicated in a similar color code as in Fig.~1 for more clarity.} 
	\label{fig:LNT1x11}
\end{figure*}

Figure \ref{fig:experimental_geometry}(b) shows a sketch of a possible outcome of a poling experiment of a multi-domain crystal after applying a positive voltage to one of the electrodes. It is well established that domain nucleation in congruent crystals of the LNO-LTO material family only starts at the +z-direction of the crystal \cite{Mackwitz2016,Sanna17surf,Younesi2021,Stanicki2020,Rusing2019SHG}. This means that inverted domains first nucleate at +z electrodes and grow towards the opposing electrode in the -z direction. Hence, four different cases in the relation of the as-grown domain pattern and the applied electric field can be distinguished [as highlighted in Fig.~\ref{fig:experimental_geometry}(b)].
In \emph{Case I} the domain between two opposing electrode fingers is oriented completely opposed to the applied electric field. Here, inverted domains can readily nucleate at the +z electrode and grow over the full gap. \emph{Case II} is the opposing case, where the as-grown domain between the electrodes is already aligned with the outer electric field. Here, nothing is expected to happen. 

If the average size of the as-grown domains are in similar order of magnitude or smaller than the gap distance $g$ between the electrodes, the chance is high that there is at-least one head-to-head or tail-to-tail domain wall within the gap. Here, we can distinguish two more cases. In \emph{Case III}, the domain directly under the positive electrode is of opposing nature (similar to \emph{Case I}). In this case nucleation of new domains is possible directly at the +z electrode and inverted domains are expected to grow until these reach the tail-to-tail domain wall, where they merge with the domain of the same orientation. 

\emph{Case IV} is the most peculiar one. Here, the domain directly underneath the +z electrode, where the positive voltage is applied, is already aligned with the electric field (like in \emph{Case II}). But somewhere within the gap one domain or more of opposing direction can be found, which may or may-not extent towards the -z electrode. If the electric field between two opposing electrodes is sufficiently large, any anti-parallel aligned domain within the gap, can invert, even if it does not intersect with the electrodes. However, aside from the necessary condition of the electric field being larger than the (effective) coercive field, the growth and nucleation of inverted domains also require the exchange of (screening) charges. For a single-domain, single crystal this is usually easily achieved, when the newly formed domains are in direct contact with an electrode (e.g. \emph{Case I} or \emph{III}). However, if the crystal is a good electrical insulator, which was shown for similarly grown lithium niobate tantalate at room temperature \cite{Yakhnevych2023}, it can be expected that poling in \emph{Case IV} is largely inhibited, because no electrical contact and charge transport is possible. Nevertheless, nucleation and poling might still be possible, if the head-to-head domain walls, where the nucleation sites are expected, are conductive and have a connection to an electrode (example highlighted with red arrows). From previous works in lithium niobate and other ferroelectrics, it is well established that domain walls can be highly conductive in certain conditions, which particular applies to head-to-head domain walls \cite{Beccard2022,Beccard2023,Kirbus19,Singh2022}. Therefore, if successful poling in LNT$_x$ in a \emph{Case IV} is observed, this can be a first hint at domain wall conductivity in LNT$_x$.

To analyze the poling results Second-Harmonic microscopy (SHM) images are taken before and after each time that electrical pulses are applied to an electrode pair. SHM is sensitive to ferroelectric domain structures and/or domain walls and is widely used to visualize domain structures \cite{Mackwitz2016,berth2007depth,spychala2020nonlinear,cherifi2021shedding,Kampfe2014SHG,Hegarty2022Darkfield,Hegarty2022Anomal,Kirbus19}. 

To test the experimental setup and electrode design and find initial poling parameters for the poling geometry, we performed poling experiments on a commercially-obtained, bulk-crystal of x-cut congruent LNO. Crucially, this also provides insights on the expected contrast from surface-near domains in SHM, which is important when interpreting SHM images from 3D domain structures \cite{cherifi2021shedding,Rusing2019SHG,Hegarty2022Anomal}. In contrast to the LNT$_x$ crystal, the LNO sample is mono-domain before poling, i.e. the whole crystal is initially homogeneous with no domain walls present. Therefore, inverted domains can only grow when a positive voltage is applied to an electrode located on the +z end of a domain (compare \emph{Case I}).

Figure \ref{fig:cLN} shows an SHM image of a poling experiment on a electrode with a gap of $g = 15$~$\mu$m and a length of several 100~$\mu$m. The image was taken with the SHM focus placed surface-near, which results for the given wavelength and laser polarization in a homogeneous, strong SH signal from the surface \cite{Hegarty2022Anomal,Hegarty2022Darkfield}. In SHM the metal electrodes always appear dark, because they do not possess any second-order nonlinearity. After application of a single electric pulse of 3~ms length with an electric field of  53.3~kV/mm within the gap, domain inversion between the fingers was observed. The applied electric field is more than two times larger then the value typically cited for z-cut congruent LNO ($\approx$ 21 kV/mm) \cite{Kim2001,Gopalan1998}. Such high fields were observed to be necessary by some authors as well \cite{Gui2009} for this poling geometry. However, there are also reports, such as Stanicki et al., who could achieve domain inversion in x-cut with electric fields of just 24~kV/mm by the use of prepulses. Here, more studies are indicated to systematically investigate the poling behavior in this x-cut geometry, which also plays a key role in thin film lithium niobate based nonlinear optical devices. One possible explanation for the large observed fields could be that in standard z-cut-poling inverted domains can directly grow vertical from the electrodes, while when using surface electrodes in x-cut poling a pronounced sideways growth into the depth of the crystal (x-direction) is required. Here, the electric field decays fast in the depth of the crystal necessitating much larger initial fields at the surface. 

In our experiment in Fig.~\ref{fig:cLN} the areas with domain inversion generate a stronger SHM signal compared to the surrounding bulk. This has been observed in SHM studies of surface-near domain inversion before \cite{berth2007depth,Gui2009,Stanicki2020,Mackwitz2016} and can be explained with constructive interference of the SHM signal generated in the surface-near inverted domain and the underlying bulk \cite{Rusing2019SHG}. We detect up to four times larger SHG signal in areas with surface-near domain inversion compared to the surrounding pristine area. The coherent interaction length for the counter-propagated SHM signal is on the order of 80~nm for the given wavelength \cite{Amber2021,Rusing2019SHG}. Therefore, a domain inversion depth of at-least 80~nm or multiples of it are expected. However, from previous work on poling of x-cut bulk and thin film it was observed that domains created in a similar configuration in congruent LNO typically penetrate 0.5 to 1~$\mu$m within the gap, which yielded similar SHM images \cite{Gui2009,Stanicki2020,Berth2009}. Further studies are required to more accurately determine the inversion depth, e.g. by three dimensional piezoresponse force microscopy \cite{roeper2024depth} or forward scattering SHM, which results in a longer coherent interaction length \cite{Amber2021}. Further, it can be noticed that the newly poled domains also show an irregular, stripy pattern, which is indicative of spike domains which are not fully merged within the inverted regions. Again, similar was observed before when poling x-cut, congruent LNO with surface electrodes \cite{Gui2009,Stanicki2020,Mackwitz2016,Berth2009}. Further information about SHM in LNT, as well as the image interpretation is discussed in the supplemental material.

Figure \ref{fig:LNT1x11} shows a result on the x-cut LNT$_x$ crystal along a 250~$\mu$m long electrode with a gap of $g= 30$~$\mu$m. Here, panel (a) shows the SHM image before, (b) after application of a positive voltage to the top electrode ('forward poling') and (c) application of the inverted voltage ('reverse poling') are shown. Further, Figures (d) and (e) show close-ups of different regions from (a) and (b). In contrast, to the poled surface-near domains from Fig.~\ref{fig:cLN} the as-grown domain walls in LNT$_x$ appear with a dark contrast. This is the typical contrast for domain walls, which are orthogonal to the surface and penetrate into the crystal \cite{Hegarty2022Anomal,Hegarty2022Darkfield,Rusing2019SHG,spychala2020nonlinear}. The domain pattern before poling is composed of many domains with typical width of 10 to 20~$\mu$m along the z-axis. Here, most of the as-grown domain walls meander irregularly and randomly through the crystal forming a complex 3D domain pattern similar to previous observations \cite{Roshchupkin2023}. In particular, these as-grown domains also form head-to-head and tail-to-tail sections, which normally are not favored in an uniaxial ferroelectric like LNO or LTO. Additionally, some brighter spots and areas can be seen, which indicate shallow, surface-near as-grown domains. Subfig.~\ref{fig:LNT1x11}(d) shows a zoom-in into three areas with domain patterns of \emph{Case I}, \emph{Case III}, and \emph{Case IV}, respectively. No domain pattern in accordance with \emph{Case II} is found in this example. The suspected polarization is sketched in an overlay. The orientation of the polarization is reconstructed based on the poling and nucleation behavior observed after poling assuming only +z nucleation typical for LNO or LTO.

A positive poling pulse was applied with a duration of 3~ms and an electric field of 50~kV/mm similar to the values determined for bulk, congruent LNO. The results are shown in Fig.~\ref{fig:LNT1x11}(b). Here, signatures of new surface-near domains can be seen for most electrode finger pairs. In more detail, it can be seen that the expected different behaviors for \emph{Cases I}, \emph{III}, and \emph{IV} can be seen as highlighted in the close-ups in Figure~\ref{fig:cLN}(e). Here, as expected  nucleation and growth is possible, when the original domain within the complete gap is oriented opposing the original domain (\emph{Case I} and \emph{Case III}).

\emph{Case IV} is the most peculiar. Here, it can be observed that nucleation at an as-grown head-to-head domain wall within the gap is possible in some areas, which is highlighted by red arrows in Fig.~\ref{fig:LNT1x11}(e). However, as discussed below and presented in the supplement (e.g. Fig.~S2) nucleation at head-to-head domain walls within the gap is not always possible. This immediately shows that reaching an electric field of the appropriate magnitude is a \emph{necessary} condition, but not a \emph{sufficient} condition to achieve domain inversion \cite{Sturman2022}. A potential mechanism allowing domain inversion in \emph{Case IV} in Fig.~\ref{fig:LNT1x11} is an enhanced conductivity of the as-grown domain walls, which can facilitate the charge exchange required for domain reversal. For the as-grown domain wall highlighted by red arrows in Fig.~\ref{fig:LNT1x11}(e) \emph{Case IV} this might be given, because this head-to-head domain wall intercepts a neighboring electrode. In this regard, the random domain pattern of the as-grown crystals offer a distinct model system to study the role of conductivity for the growth and nucleation of domains. However, further investigations, e.g. by conductivity atomic force microscopy \cite{Singh2022} is necessary to confirm the domain wall conductivity. For future work, it may be possible to locate regions, which are similar to \emph{Case I}, \emph{before} electrode fabrication, e.g. with SHM, to obtain the best poling results. Further, it is noticed that in a few locations domain inversions also happen in \emph{Case IV} scenarios, such as next to the scale bar in Fig. \ref{fig:LNT1x11}(b). Here, these inversions might indicate buried (conductive) domain features, or local defects or stoichiometry variations which lower the coercive field or yield sources of screening charges. Here, more detailed investigations, e.g. by SHM, 3D piezo-response force microscopy or Raman spectroscopy \cite{roeper2024depth,Reitzig2021,Hegarty2022Anomal}, are required to analyze (and possibly prohibit) such inversions.

When applying a reversed voltage to the electrode pair domain growth can now start from the opposing electrodes, if the appropriate original domain orientation is given, as depicted in Fig.~\ref{fig:LNT1x11}(c). Some back poling, i.e. a reversal of the previously poled domains, happens, but many poled domains from before (b) stay stable and are not influenced by a reverse poling pulse. This indicates the strongly pinned domains in the LNT$_x$ crystal.

In the supplement, three more examples from the LNT$_x$ crystal are shown. Here, Fig.~S1 shows a poling result on a electrode pair, which features a line defect running in the middle of the gap across most of the electrode length, which acts as a pinning site stopping domain growth from both sides. Due to the strong pinning by this defect structure, a periodic structure can be achieved after a positive pulse was applied to both electrodes with only minimal reverse poling. 

In contrast, Fig.~S2 depicts an electrode pair with a particular high density of as-grown domains with average distances of 5~$\mu$m or less. Here, even though that fields up to 60~kV/mm are applied, only very few electrodes show signs of nucleation. If new domains nucleate, they immediately intercept a domain of the same orientation. Almost no additional nucleation of new domains within the gap similar to \emph{Case IV} is observed. However, in this electrode pair, very few head-to-head domain walls, which could act as potential nucleation sites, appear to be connected with an electrode. This also supports the fact that the domain wall in Fig.~\ref{fig:LNT1x11} \emph{Case IV} might be conductive and in contact with the electrodes to allow for nucleation, suggesting further study.

Fig.~S3 depicts a case, where parts of the electrodes are consistent with a \emph{Case II} after the first pulse and behave like a \emph{Case I} in reverse poling.

In conclusion, in this work we have investigated surface-near (periodic) domain inversion over a poling gap of up to 30~$\mu$m with a period of 10~$\mu$m is possible within mixed crystals of lithium niobate tantalate. In this regard, surface-near, local poling presents a possible strategy to achieve periodic poling even in multi-domain crystals, if a gap size smaller than the average domain width is chosen. In the future, this technique can be combined with large-scale domain sensitive imaging techniques, like SHM \cite{Hegarty2022Anomal,Hegarty2022Darkfield,Kampfe2014SHG}, to select suitable areas before electrode fabrication. This lays the foundation to create nonlinear optical devices or tailor conductive domain walls within the whole LNT$_x$ material family. The poling of the as-grown domain structures provide an intricate model systems how both, defects and existing domain structures, can inhibit and pin domain growth on a larger scale within the lithium niobate tantalate material systems.

\section*{Supplementary Material}
In the supplementary material three more examples of poled structures in the LNT mixed crystal with different as-grwon domain sizes are shown. Additionally, relevant fundamentals of SHM and the interpretation of SHM images in LNT are explained.

\begin{acknowledgments}The authors gratefully acknowledge financial support by the Deutsche Forschungsgemeinschaft (DFG) through collaborative research centers SFB-TRR142, project B07, (ID: 231447078) and CRC1415 (ID: 417590517), the Forschungsgruppe FOR5044 (ID: 426703838; \url{http:\\www.For5044.de}), the Würzburg-Dresden Cluster of Excellence ”ct.qmat” (EXC 2147), as well as by the Federal Ministry of Education and Research / Bundesministerium für Bildung und Forschung (Grant number 13N15975). We thank Dr. Steffen Ganschow from the IKZ Berlin for providing the crystal; and Thomas Gemming and Dina Bieberstein for assistance in dicing.
\end{acknowledgments}

\section*{References} 
\bibliography{LNT_poling}

\beginsupplement

\clearpage

\onecolumngrid

\section*{Supplement}

\subsection*{Further examples of poling in LNT}

\begin{figure*}[h]
	\centering
	\includegraphics[width=0.7\linewidth]{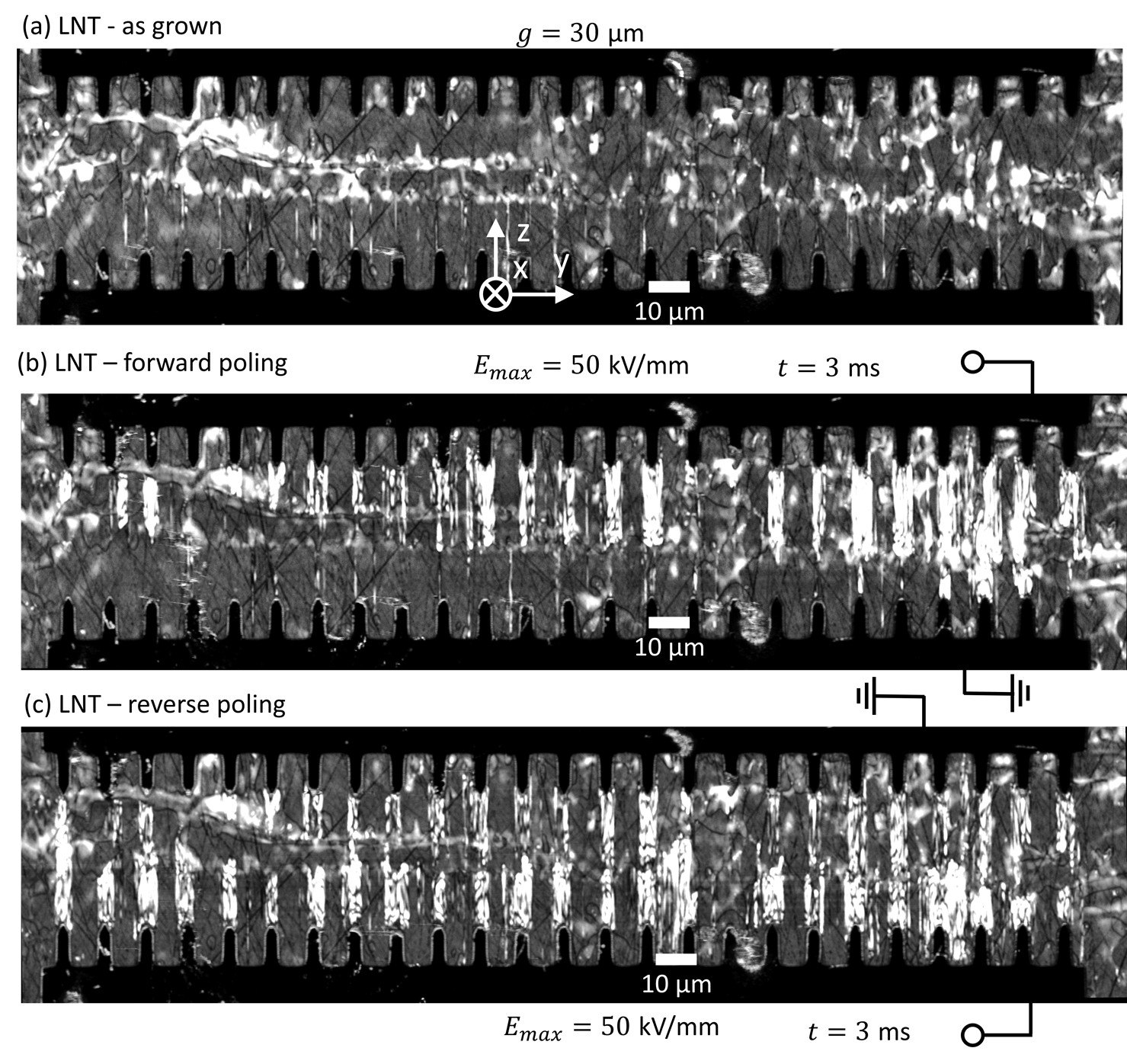}
	\caption{Second harmonic microscope image of a 250~$\mu$m electrode pair on x-cut lithium niobate tantalate mixed crystals (a) before poling, (b) after application of a positive voltage pulse with an electric field of $E_{max} = 50$~kV/mm in the gap with $g=20$~ $\mu$m and a duration of $t=3$~ms and (c) after application of the same electric pulse with reversed polarity. (a) Before poling, a random domain structure with meandering domains of 10 to 20~$\mu$m diameter is visible. (b) After forward poling signatures for surface-near, shallow domains with a depth of $<1$~ $\mu$m can be seen. An as-grown domain wall apparently runs across the electrode gap, which acts as a stop for domain growth. (c) After application of a reverse pulse, inverted domains also nucleate at the bottom electrodes. Due to apparent strong pinning only little back poling is observed and a quasi-periodic structure can be achieved.} 
	\label{fig:LNT1x10}
\end{figure*}

\begin{figure*}[h]
	\centering
	\includegraphics[width=0.7\linewidth]{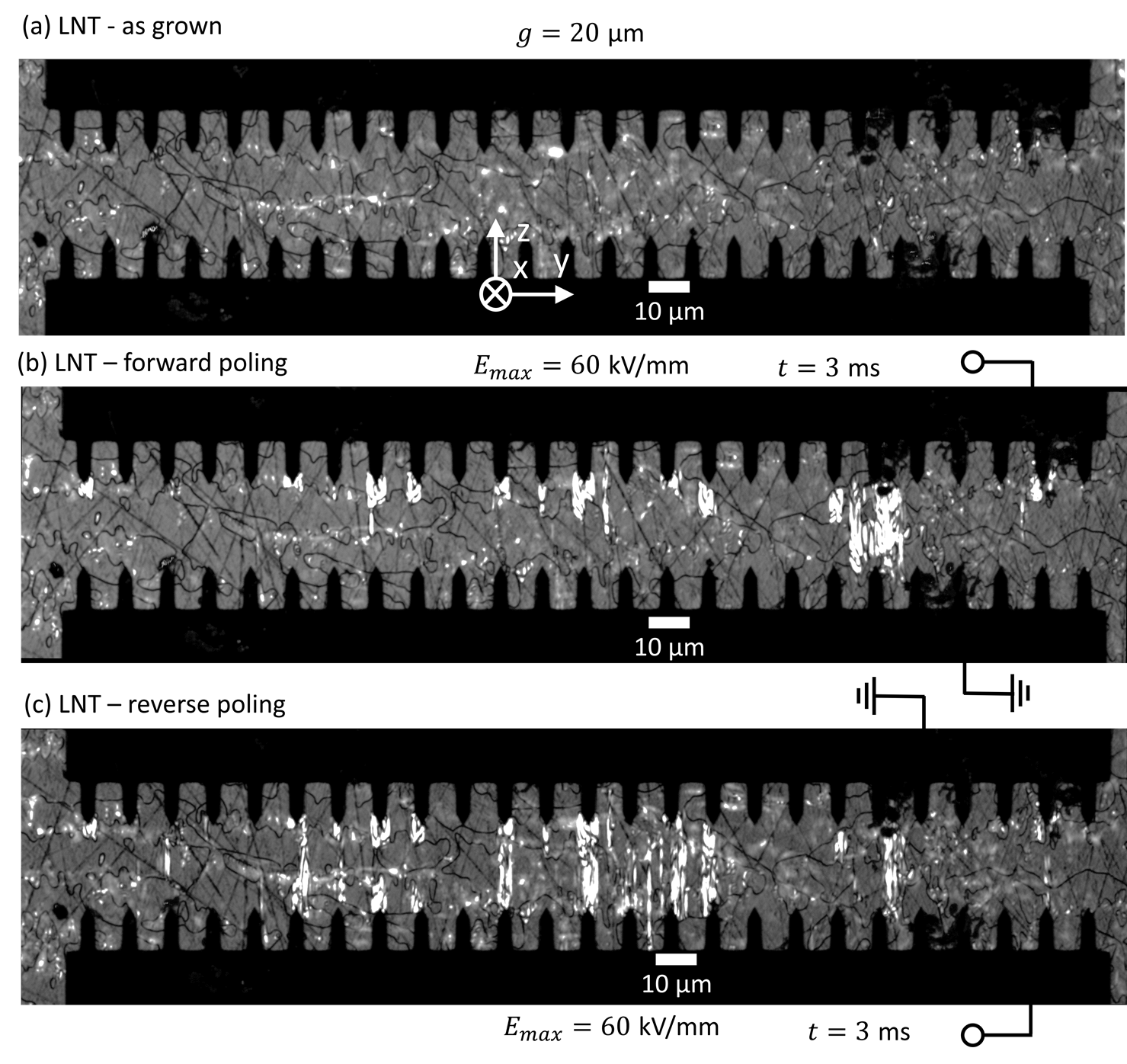}
	\caption{Second harmonic microscope image of a 250~$\mu$m electrode pair on x-cut lithium niobate tantalate mixed crystals (a) before poling, (b) after application of a positive voltage pulse with an electric field of $E_{max} = 60$~kV/mm in the gap with $g=20$~$\mu$m and a duration of $t=3$~ms and (c) after application of the same electric pulse with reversed polarity. (a) Before poling, a random domain structure with meandering domains of only 5 to 10~$\mu$m sizes are visible, which are smaller than in the other three presented examples. (b) Even though the electric field in this experiment is even higher, nucleation is only observed at a few electrode fingers indicating that the much smaller as-grown domain structure compared to the other experiments inhibits domain nucleation and growth. (c) Even when a reverse voltage is applied, only few electrodes show nucleation of inverted domains.} 
	\label{fig:LNT2x8}
\end{figure*}

\begin{figure*}[h]
	\centering
	\includegraphics[width=0.7\linewidth]{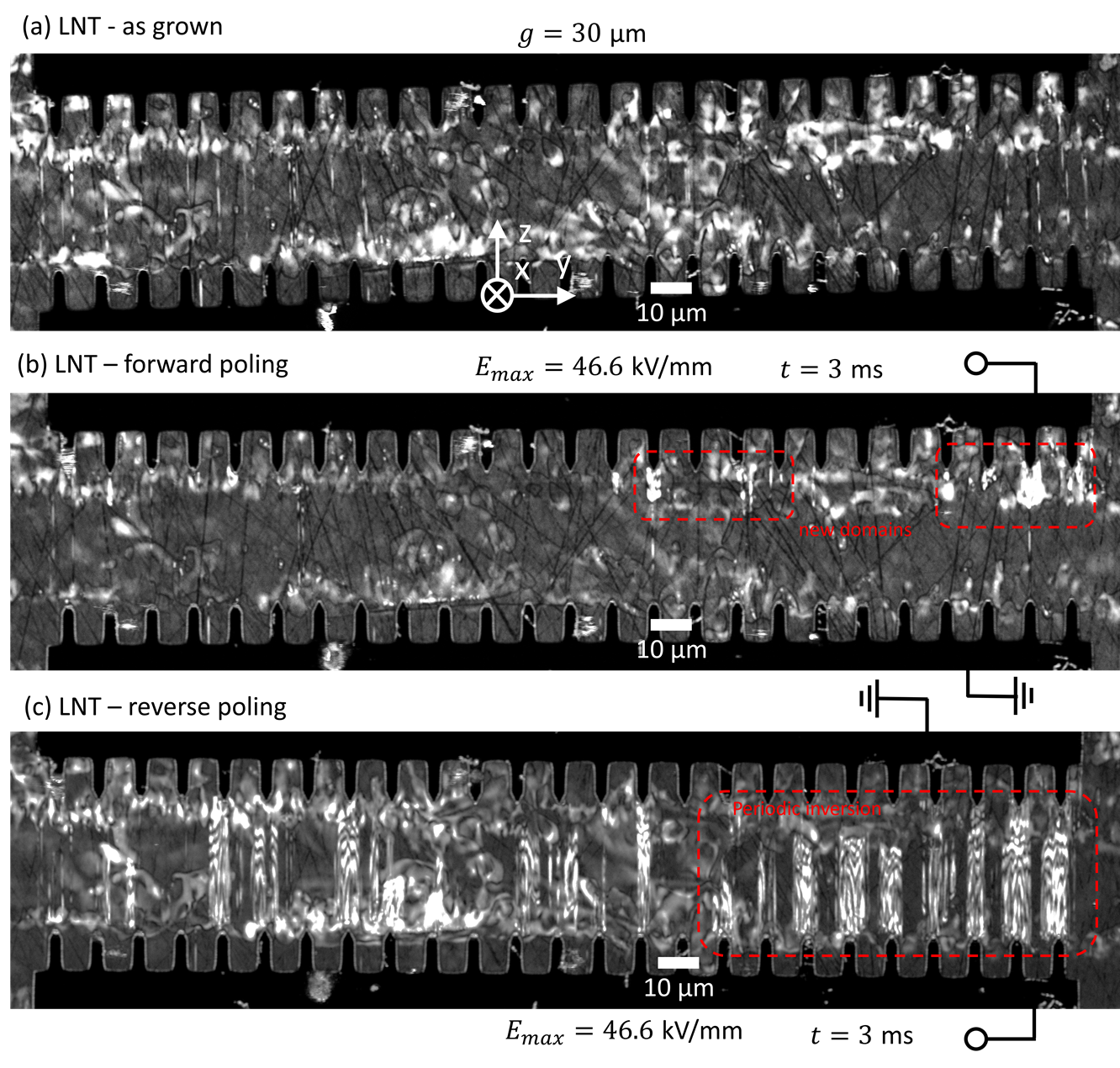}
	\caption{Second harmonic microscope image of a 250~$\mu$m-long electrode pair on x-cut lithium niobate tantalate mixed crystals (a) before poling, (b) after application of a positive voltage pulse with an electric field of $E_{max} = 46.6$~kV/mm in the gap with $g=30$~$\mu$m and a duration of $t=3$~ms and (c) after application of the same electric pulse with reversed polarity. (a) Before poling, almost no domain walls, especially in the bottom right part of the image, are visible, but only some signatures for natural occuring shallow domains. (b) After an application of an initial positive voltage to the top electrode, almost no poling is observed, as most of the area between the electrodes are described by \emph{Case II}, while on some places newly formed domains immediately intercept DWs. (c) In contrast, after reverse poling, especially in the bottom right area a periodic structure forms with domains penetrating the full gap length. However, in the left part of the electrode pair, many newly formed domains are intercepted by shallow domains.} 
	\label{fig:LNT1x12}
\end{figure*}

\subsection*{Fundamentals of SHG microscopy in LNT}

\begin{figure*}[h!]
	\centering
	\includegraphics[width=1\linewidth]{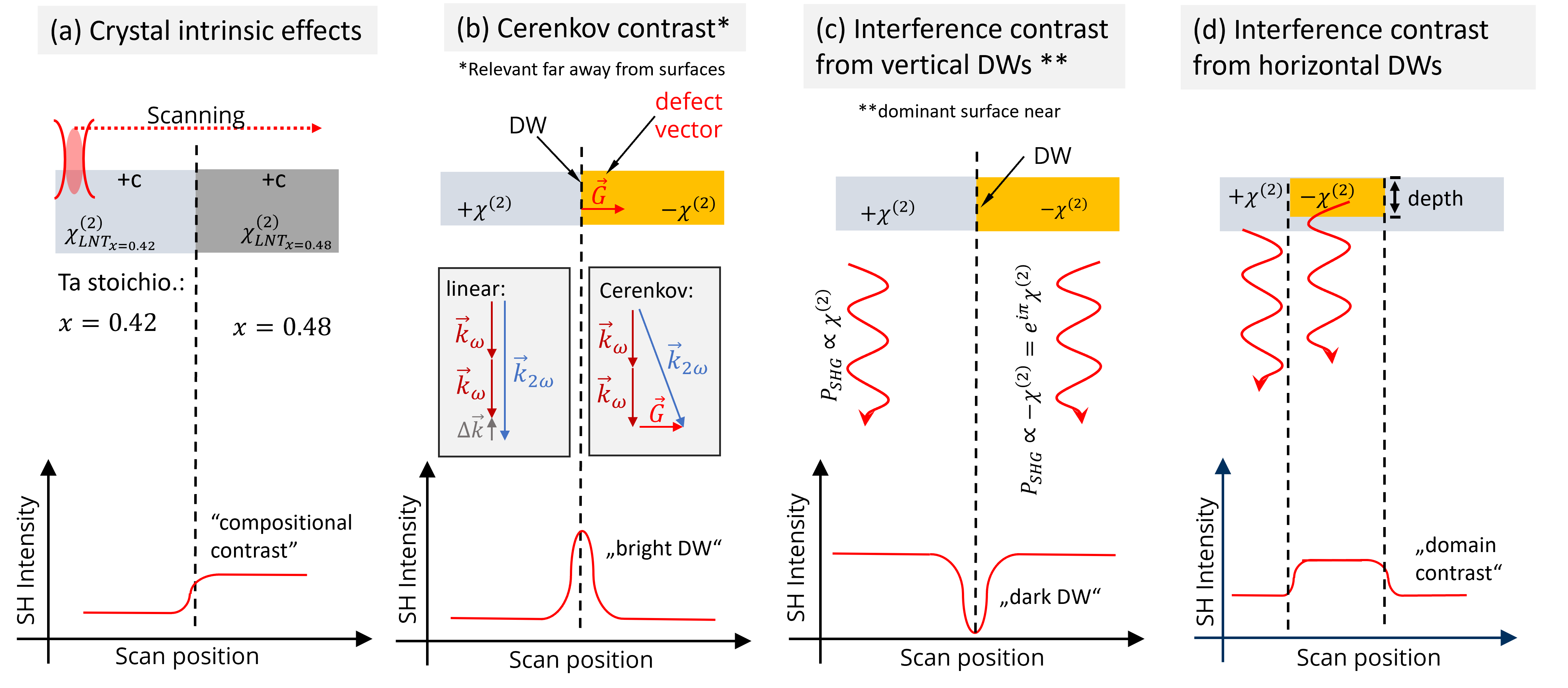}
	\caption{Relevant contrast mechanisms for SHG microscopy in a multi-domain, mixed crystal like LNT.} 
	\label{fig:Kontrast}
\end{figure*}

In this work we use Second Harmonic Generation (SHG) microscopy to image and identify the domain structures pre- and post-poling. We use this information to create false-color images of the domain structures to visualize the effect of poling in the four highlighted \emph{Cases}. SHG microscopy is a very common method to visualize ferroelectric domain structures \cite{Hegarty2022Anomal,berth2007depth,spychala2020nonlinear,Rusing2019SHG}. However, the appearance of domains or domain walls does drastically depend on the conditions of the imaging (focus depth, wavelength, type/orientation of domain wall), which can result in a bright domain wall (DW) contrast, dark DW contrast or domain polarity contrast \cite{Hegarty2022Anomal,berth2007depth,spychala2020nonlinear,Rusing2019SHG,Kaneshiro2008,Kaneshiro2010}. In a crystal system like LNT, further, local variations of the Nb-Ta ratios, which are common in Czochralski grown crystals \cite{Roshchupkin2023,Bashir2023}, might result in similar contrasts. Therefore, clear considerations most be taken, when domain imaging in a complex, not well understood system like LNT is performed.

Some potential signatures of domain walls, domains or local variations of the Nb-Ta content are highlighted in Fig.\ref{fig:Kontrast}. Fig.~\ref{fig:Kontrast}(a) shows a possible contrast, when a local Nb-Ta variation are observed. Here, a change in SHG signal might be expected between areas of different Nb-Ta ratios, which result in a change of the local, nonlinear susceptibility $\chi^{(2)}$. Similarly, local strains or defects or changes in local phasematching due to changes in refractive index might result in a contrast, too. These might be mistaken for a domain or domain wall and need to be seperated. Domain walls (in a otherwise homogeneous material) can appear as bright or dark, while in certain conditions the domain polarity might be visible as well. The most important DW contrasts relevant for this work are shown in Fig.~\ref{fig:Kontrast}(b)-(c). 

The main reason for a bright DW contrast is the Cerenkov-type phase matching, see Fig.~\ref{fig:Kontrast}(b). In Cerenkov-type phase matching, the DW can emit more light because the DW provides an additional phase matching vector, which improves the phase matching between the fundamental and SHG beam \cite{Hegarty2022Anomal,Hegarty2022Darkfield}. This contrast is particular dominant, if SHG in the bulk, i.e. in the absence of a DW is forbidden or much weaker compared to the Cerenkov-type signal. This is usually the case if imaging is performed for normal dispersion \cite{Hegarty2022Anomal} (refractive index at the SHG wavelength is larger compared to the refractive index at the fundamental wavelength) \emph{and} 
the focus is placed away from the surface of the crystal, i.e. at least a few Rayleigh ranges below the crystal's surface \cite{spychala2020nonlinear}. Here, SHG within a homogeneous domain (no domain walls or other changes) is forbidden due to the so called Guoy-phase shift in the focus. Therefore, DWs do appear bright, as the Cerenkov-type phase matching results in more signal compared to the surrounding (single-domain) bulk. Please note, in certain conditions other contrasts can result in bright DWs as well, but these effects are usually weaker than Cerenkov-type contrast and are usually limited to the surface of the crystal \cite{spychala2020nonlinear,cherifi2021shedding}.

If the focus is placed close to the surface, which is the case for most of the images of this work, the Cerenkov-contrast becomes less relevant, as SHG light from a single domain (in the absence of DWs) is generated \cite{spychala2020nonlinear,Hegarty2022Anomal,Hegarty2022Darkfield}. Here, (vertical) DWs usually appear dark due to interference of the SHG light, which is generated on either side of the the DW [Fig.~\ref{fig:Kontrast}(c)]. Here, from the viewpoint of SHG the main difference between domains of opposing direction is the sign of the nonlinear susceptibility, which is inverted between domains of opposing polarity. This results in an extra $\pi$-phase shift between the SHG light generated between neighboring domains. If the focus is now placed exactly on top of the domain, the light generated in one domain will exactly cancel with light in the opposing domain, which results in a dark domain wall. As the bulk SHG effect is usually stronger that the Cerenkov-type phase matching, this interference contrast will dominate at the surface \cite{Hegarty2022Darkfield,spychala2020nonlinear}.

\begin{figure*}[h!]
	\centering
	\includegraphics[width=1\linewidth]{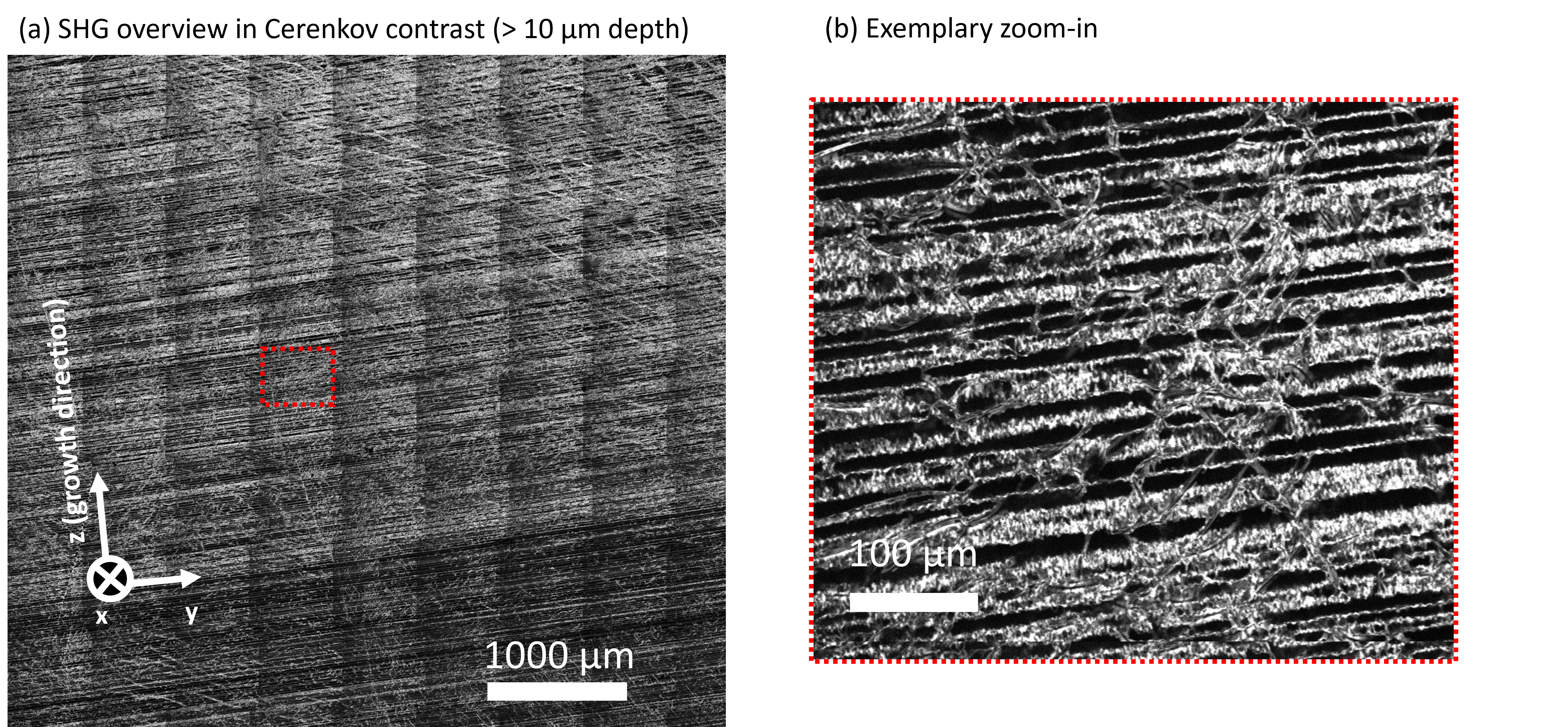}
	\caption{Second harmonic microscope image of another x-cut wafer cut from the same boule as the crystal used for poling. Here, layered domains with head-to-head and tail-to-tail configurations can be observed. This image was taken in a depth of $> 10 \mu$m, which leads to a dominating Cerenkov-type contrast. The image contains a visible "tiling", which is a result of stitching of the SHG microscopy image from individual frames. The image is stitched from frames of $\approx 500 \times 600 \mu$m$^2$, which is the maximum area that could be imaged with the given objective lens without moving the sample. (b) The zoom-in shows an even more complex domain structure with some areas of up to $20 \mu$m being single domain. Please note, this type of SHG microscopy is sensitive to the domain walls, but not the domain orientation. Therefore, the orientation ($+P_S$ vs $-P_S$) cannot be inferred from the images, but can be deduced after poling.} 
	\label{fig:LNT45overview}
\end{figure*}

If the domain wall is vertical and the domain is shallow, i.e. its depths is only on the order of a micron, this can also result in a strong interference contrast (constructive or destructive) [Fig.~\ref{fig:Kontrast}(d)]. In this work, we observe this contrast for the newly poled domains in the reference crystal of LN, as well as for LNT. Here, the SHG signal in this shallow domain is stronger compared to the surrounding. Please note, this type of interference contrast can be both constructive or destructive, which depends on the coherent interaction length and depth of the domain. In this work we measure in back-reflection, i.e. through the same objective lens as the excitation. In backwards direction the coherent interaction length $l_c$ is on the order of 80~nm for the given wavelength \cite{Rusing2019SHG}. Therefore, the SHG signal will show a maximum if the surface-near domain has a depth of at least 80~nm. However, if the domain penetrates deeper, further maxima of the SHG signal will be observed when the domain has depths of multiples of $m$ with  $2l_c (m + 1)/2$. Therefore, the domains have a depth of atleast 80~nm. 
This is consistent with previous reports of poling of x-cut bulk and thin films in LNO, where TFLN films of 600 to 800 nm thickness can be fully poled with surface electrodes \cite{Zhao2023,Reitzig2021,Stanicki2020}. To further analyze the poling depth, measurements in forward scattering, where the coherent interaction length is more than 1 $\mu$m or destructive measurements, such as etching of trenches, are required.

\subsection*{Domain structures in Czochralski-grown LNT crystals}

Prior to poling the orientation of the natural domain structures in the as-grown crystals were analyzed with SHG microscopy. Figure \ref{fig:LNT45overview} shows a large scale SHG image of an x-cut wafer cut from the same boule as the wafer that was used for poling. This piece is cut next to the wafer used for poling and provides insight into the expected domain configuration in the LNT crystal. The SHG microscopy image was taken in a depth of more than 10~$\mu$m, where the Cerenkov contrast dominates. In the $\approx 5 \times 5$ mm$^2$ overview many horizontal lines parallel to the y-crystal direction can be seen. These bright lines are signatures of domain walls, which apparently orient themselves in a head-to-head and tail-to-tail configuration, which is usually not observed in LNT mixed crystals. These domain walls are oriented orthogonally to the pulling direction, which is along the z-axis. From studies of the crystal growth conditions of LN-LT solid solutions it is well known that the large separation of the liquidus and solidus line (in particular for crystals with Ta concentrations far away from the end compounds ($0.9 > x > 0.1$) will lead to slight variations in Ta-Nb content orthogonal to the pulling direction. Here, layers of slightly larger and slightly lower Ta content will form \cite{Roshchupkin2023,Bashir2023}. It can be suspected that these layers serve as pinning boundaries for the layered head-to-head and tail-to-tail domain boundaries observed in the crystals. However, here, further studies with different methods are required beyond the scope of this work.

Figure \ref{fig:LNT45overview}(b) shows a zoom-in an even more complex domain structure with more domain walls penetrating into the depth of the picture, as well as running parallel to the z-axis. However, some areas an be indentified, which appear to be single domain. This type of SHG microscopy is sensitive to the domain walls, but not the domain orientation. Therefore, the orientation ($+P_S$ vs $-P_S$) cannot be inferred from the images, but can be inferred after poling.

\begin{figure*}[h!]
	\centering
	\includegraphics[width=1\linewidth]{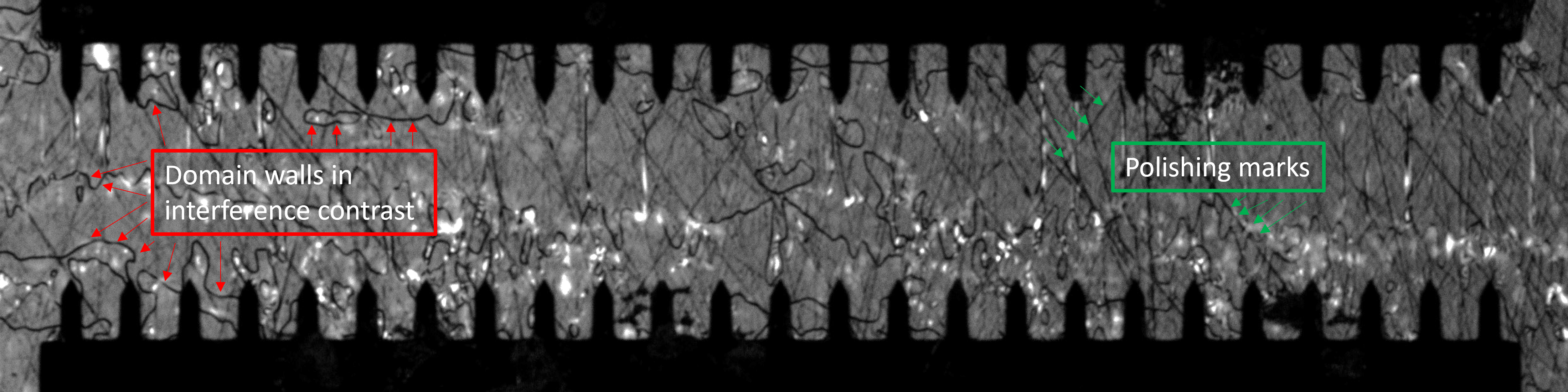}
	\caption{Example SHG picture taken with the focus of the SHG microscope placed at the surface of the sample. Here, the interference contrast will dominate, i.e. domain walls will appear dark. It can be seen that the DW form irregular shapes and lines that meander between the electrodes. Please note, shallow domains, which are also naturally appear in the samples, will appear bright due to enhance phase matching. In contrast, straight black lines are caused by nanoscopic scratches as a leftover of incomplete polishing.} 
	\label{fig:example}
\end{figure*}

Figure \ref{fig:example} shows an examplary picture of an electrode taken, when the focus is placed surface near. In this case, the interference contrast [see Fig.~\ref{fig:Kontrast}(c)] will be the dominating effect and DWs will appear dark. Some examplary domain walls are highlighted in the figure to demonstrate how the pictures are interpreted and analyzed. 

\newpage

\end{document}